\def\cm{{\rm\thinspace cm}}

\def\erg{{\rm\thinspace erg}}

\def\K{{\rm\thinspace K}}
\def\keV{{\rm\thinspace keV}}
\def\km{{\rm\thinspace km}}

\def\Mpc{{\rm\thinspace Mpc}}
\def\Msun{\hbox{$\rm\thinspace M_{\odot}$}}

\def\s{{\rm\thinspace s}}
\def\yr{{\rm\thinspace yr}}

\def\pcmcu{\hbox{$\cm^{-3}\,$}}

\def\ergps{\hbox{$\erg\s^{-1}\,$}}

\def\kmps{\hbox{$\km\s^{-1}\,$}}

\def\Msunpyr{\hbox{$\Msun\yr^{-1}\,$}}

\def\pcmsq{\hbox{$\cm^{-2}\,$}}

\def\mdot{\hbox{$\dot{m}$}}
\def\Mdot{\hbox{$\dot{M}$}}
\def\spose#1{\hbox to 0pt{#1\hss}}
\def\approxlt{\mathrel{\spose{\lower 3pt\hbox{$\sim$}}
        \raise 2.0pt\hbox{$<$}}}
\def\approxgt{\mathrel{\spose{\lower 3pt\hbox{$\sim$}}
        \raise 2.0pt\hbox{$>$}}}

\documentstyle[psfig]{mn}

\title{The ``quiescent'' black hole in M87}

\author[C.~S.~Reynolds et al.]
{C.~S.~Reynolds$^1$\thanks{Present address: JILA, Campus Box 440,
University of Colorado, Boulder CO 80309, USA}, T.~Di~Matteo$^1$, A.~C.~Fabian$^1$, U.~Hwang$^2$ and
C.~R.~Canizares$^3$\\ 
{\small $^1$Institute of Astronomy, Madingley Road,
Cambridge, CB3 OHA}\\ 
{\small $^2$NASA Goddard Space Flight Center,
Greenbelt, MD 20771, USA}\\ 
{\small $^3$Center for Space Research,
Massachusetts Institute of Technology, Massachusetts Avenue,
Cambridge, MA 02139, USA} }

\begin{document}

\maketitle

\begin{abstract}
It is believed that most giant elliptical galaxies possess nuclear
black holes with masses in excess of $10^8\Msun$.  Bondi accretion
from the interstellar medium might then be expected to produce
quasar-like luminosities from the nuclei of even quiescent elliptical
galaxies.  It is a puzzle that such luminosities are {\it not
observed}.  Motivated by this problem, Fabian \& Rees have recently
suggested that the final stages of accretion in these objects occurs
in an advection-dominated mode with a correspondingly small radiative
efficiency.  Despite possessing a long-known active nucleus and
dynamical evidence for a black hole, the low radiative and kinetic
luminosities of the core of M87 provide the best illustration of this
problem.  We examine an advection-dominated model for the nucleus of
M87 and show that accretion at the Bondi rate is compatible with the
best known estimates for the core flux from radio through to X-ray
wavelengths.  The success of this model prompts us to propose that
FR-I radio galaxies and quiescent elliptical galaxies accrete in an
advection dominated mode whereas FR-II type radio-loud nuclei possess
radiatively efficient thin accretion disks.
\end{abstract}


\section{Introduction}

There is strong evidence that most giant elliptical galaxies should
possess nuclear supermassive black holes, relics of an earlier quasar
phase.  Quasar counts and integrated luminosities suggest masses above
$10^8$ -- $10^9\Msun$.  Given this fact, there is a major puzzle
surrounding quiescent giant ellipticals which was first illuminated by
Fabian \& Canizares (1988).  A massive black hole in the centre of an
elliptical galaxy would accrete from the interstellar medium (which
forms a hot hydrostatic atmosphere in the potential well of the
galaxy).  The accretion rate would be expected to be at least that
given by Bondi's spherical accretion formula (Bondi 1952).  If the
resulting accretion flow into the hole proceeds via a standard
accretion disk (with a radiative efficiency of $\sim 10$ per cent),
all such nuclei should be seen to possess quasar-like luminosities.
This is contrary to observation.

The nearby giant elliptical galaxy M87 (NGC~4486) might be considered
a counter example because it has long been known to host an active
nucleus that powers a jet and the giant radio lobes of Virgo A.
Furthermore,   {\it HST} observations have now provided a
direct dynamical determination of the nuclear black hole mass of
$M\approx 3\times 10^9\Msun$ (Ford et al. 1995; Harms et al. 1995).
In fact, M87 illustrates the problem of quiescent black holes in giant
ellipticals and, we suggest, illuminates the solution.  Qualitative
evidence for the relative quiescence of M87 comes from a comparison to
the quasar 3C273, which presumably contains a black
hole of comparable mass. While both have core, jet and lobe emission,
the luminosity of M87 in all wavebands falls 5 orders of magnitude
below that of 3C273 (see below).

 The contrast between M87 and 3C273 cannot be completely ascribed to
a smaller mass accretion rate in the former, as can be seen by an
estimate of the Bondi accretion rate in M87. Imaging X-ray
observations provide information on the hot
interstellar medium (ISM).  A deprojection analysis of data from the
{\it ROSAT} High Resolution Imager (HRI) shows that the ISM has a
central density $n\approx 0.5\pcmcu$ and sound speed
$c_{\rm s}=500\kmps$ (C.~B.~Peres, private communication).  The
resulting Bondi accretion rate onto the central black hole is
$\Mdot\sim 0.15\Msunpyr$.  Following standard practice, we define a
dimensionless mass accretion rate by
\begin{equation}
\mdot=\frac{\Mdot}{\Mdot_{\rm Edd}},
\end{equation}
where $\Mdot$ is the mass accretion rate and $\Mdot_{\rm Edd}$ is the
Eddington accretion rate assuming a radiative efficiency of
$\eta=0.1$.  For M87, the Eddington limit is $L_{\rm Edd}\approx
4\times 10^{47}\ergps$ corresponding to $\Mdot_{\rm Edd}=65\Msunpyr$.
The Bondi accretion rate corresponds to $\mdot\sim 2\times 10^{-3}$
and so, assuming a radiative efficiency $\eta=0.1$, would produce a
radiative luminosity of $L\sim 8\times 10^{44}\ergps$.
Observationally, the nucleus is orders of magnitude less active.  The
observed radiative power does not exceed $L_{\rm obs}\sim
10^{42}\ergps$ (Biretta, Stern \& Harris 1991; also see Section 2 of
this letter) and the time-averaged kinetic luminosity of the jet
cannot exceed much more than $L_{\rm K}\sim 10^{43}\ergps$ (Reynolds
et al. 1996).

The recent interest in advection-dominated accretion disks (Narayan \&
Yi 1995; Abramowicz et al. 1995; Narayan, Yi \& Mahadevan 1995)
prompted Fabian \& Rees (1995) to suggest that such disks exist around
the nuclear black holes in quiescent giant elliptical galaxies.  In
this mode of accretion, the accretion flow is very tenuous and so a
poor radiator. (The possibility of similarly tenuous `ion-supported
tori' had been discussed in the context of radio galaxies by Rees et
al. 1982 and for the Galactic centre by Rees 1982).  Much of the
energy of the accretion flow cannot be radiated and is carried through
the event horizon.  Fabian \& Rees (see also Mahadevan 1996) realised
that the resulting low radiative efficiency provides a possible
solution to the elliptical galaxy problem described above.  They
identify the weak parsec-scale radio cores seen in most elliptical
galaxies (Sadler et al. 1989; Wrobel \& Heeschen 1991; Slee et
al. 1994) with synchrotron emission from the plasma of the
advection-dominated disks (ADD).

In this {\it Letter} we present a detailed examination of the
possibility that the massive black hole in M87 accretes via an ADD.
In particular, we compute the spectrum of the ADD and show that it is
consistent with the observations for physically reasonable mass
accretion rates.  In Section 2 we compile data from the literature on
the full-band spectrum of the core of M87 and present some additional
data on the X-ray flux from the core.  Care is taken to limit the
effect of contaminating emission from the jet and/or the galaxy.
Despite this, the spectrum we obtain must still be considered as a set
of upper limits on the spectrum of the accretion flow with the
non-thermal emission from the jet representing the main contaminant.
We make a direct comparison to the quasar 3C~273.
Section 3 describes some details of our ADD model spectrum
calculation.  Section 4 compares this model spectrum with the data and
finds that accretion rates comparable with the Bondi rate do not
overproduce radiation and are thus acceptable.  Section 5 discusses
some further astrophysical implications of this result.

\section{The spectrum of the core emission}

\subsection{The M87 data}

\begin{table*}
\caption{Summary of data for the core of M87.}
\begin{center}
\begin{tabular}{ccccc}\hline
Frequency & Resolution & $\nu F_{\nu}$ & reference & notes \\
$\nu$ (Hz) & (milliarcsecs) & (10$^{-14}$\,erg\,s$^{-1}$\,cm$^{-2}$) & & \\\hline
$1.7\times 10^9$ & 5 & 1.65 & Reid et al. (1989) & VLBI \\
$5.0\times 10^9$ & 0.7 & 1.0 & Pauliny-Toth et al. (1981) & VLBI\\
$2.2\times 10^{10}$ & 0.15 & 4.8 & Spencer \& Junor (1986) & VLBI\\
$1.0\times 10^{11}$ & 0.1 & 8.7 & B{\"a}{\"a}th et al. (1992) & VLBI\\
$7\times 10^{14}$ & 50 & 200 & Harms et al. (1994) & HST \\
$2.4\times 10^{17}$ & $4\times 10^3$ & 85 & Biretta et al. (1991) & {\it Einstein} HRI\\
$2.4\times 10^{17}$ & $4\times 10^3$ & 160 & this work & {\it ROSAT} HRI\\
$4.8\times 10^{17}$ & $2\times 10^5$ & $\le700$ & this work & {\it ASCA} \\\hline
\end{tabular}
\end{center}
\end{table*}

 \begin{figure}
 \centerline{\psfig{figure=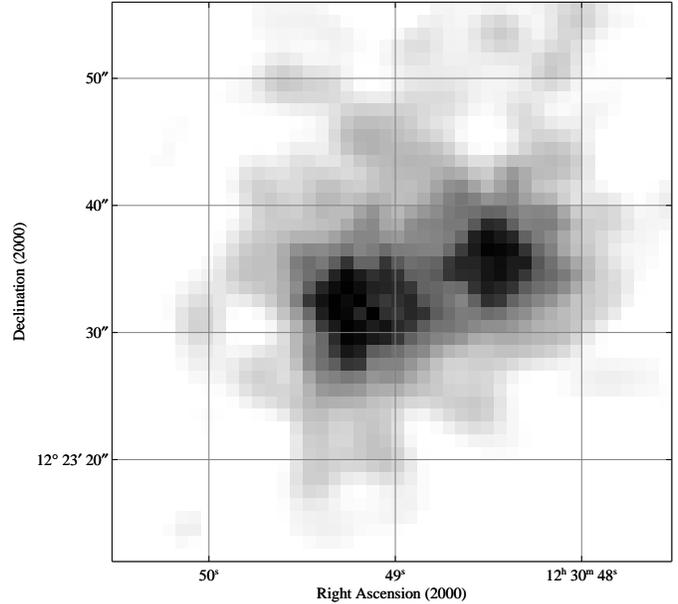,width=0.5\textwidth}}
 \caption{The core regions of M87/Virgo as imaged in a 14ks exposure with 
 the {\it ROSAT} HRI.  Two distinct sources are seen embedded in
 general diffuse emission.  The easternmost source corresponds to the
 core of M87 whereas the western source coincides with the
 brightest knot within the jet (knot-A).  The diffuse emission is from
 the hot interstellar medium.}
 \end{figure}

\noindent In order to examine the nature of the accretion flow in M87, we have
attempted to compile the best observational limits on the full band
spectrum of the core emission.  Our aim is to obtain good
observational limits on the core flux over a wide range of frequencies
rather than to compile a comprehensive list of all previous
observations.  For radio through optical, we use the highest spatial
resolution data available from the literature in order to minimize the
contribution to the flux from the synchrotron emitting jet and the
galaxy.  However, contributions from the jet and the underlying galaxy
are unavoidable and so the derived spectrum should be considered an
upper limit to that of the accretion flow at the core of M87. These
data are summarized in Table~1.

The situation with regard to hard X-ray emission from the AGN in M87
has been confused and deserves some comment.  The overall X-ray
emission from M87 is highly extended and clearly of thermal origin,
arising from diffuse gas with average kT$\sim3$ \keV.  Several
observations with wide-field X-ray instruments indicated the presence
of an X-ray power law component, which was suggested to be associated
with the nuclear region of M87 (Lea et al.  1981, Matsumoto et
al. 1996 and references therein), but were probably confused by the
hard X-ray emission from the nearby Seyfert 2 galaxy NGC~4388 (Hansen
1990; Takano \& Koyama 1991).  A reanalysis of soft X-ray {\it
Einstein} HRI observations with arc second resolution by Biretta,
Stern \& Harris (1991) found emission from a core point source at a
level more than an order of magnitude below that found by the
wide-field X-ray instruments.

We have examined the {\it ROSAT HRI} and ASCA data sets in order to
constrain further the nuclear X-ray flux of M87.  The {\it ROSAT} data
were retrieved from the public archive situated at the {\sc goddard
space flight center} and result from a 14\,200\,s exposure performed
on 1992-June-7.  Figure~1 shows the {\it ROSAT} HRI image of the
central regions of M87.  Emission from the core and knot-A ($\sim
10$\,arcsecs west of the core) are clearly separated from each other
as they were in the {\it Einstein} HRI image (Biretta, Stern \& Harris
1991).  The {\it ROSAT} HRI count rate is $0.093$\,cts\,s$^{-1}$
(determined using the {\sc ximage} software package).  Assuming the
spectrum to be a power-law with a canonical photon index $\Gamma=1.7$
(Mushotzky, Done \& Pounds 1993) modified by the effects of Galactic
absorption (with column density $N_{\rm H}=2.5\times 10^{20}\pcmsq$;
Stark et al. 1992), this count rate implies a flux density at 1\,keV
of $F(1\,{\rm keV})=0.67\,\mu$Jy.  This result is fairly insensitive
to choosing a different power-law index.

ASCA observed M87 during the PV phase: 12\,600\,s of good data were
obtained in 1993 June.  Independent analysis was performed by
Matsumoto et al. (1996).  A 3\,arcmin radius circle centered on the
nucleus of M87 contains approximately 50\,000 counts in a single SIS
detector. We performed a variety of fits to the spectrum in the
central regions, incorporating multiple thermal components, possible
excess low-energy absorption, and possible cooling-flow emission. In
no case does the addition of a power law component give a noticeable
improvement in the fit (we fix the power law index at
$\Gamma=1.7$). Our limit (at 90 per cent confidence) is listed in
Table 1. It is somewhat below the value (whose significance is not
stated) reported by Matsumoto et al. (1996).  The {\it Einstein} and
{\it ROSAT} HRI fluxes differ somewhat.  Given the well known
variability of AGN in the X-ray band (Mushotzky, Done \& Pounds 1993),
time variability is a plausible explanation for this difference.

High resolution VLBI observations probably provide the strongest
constraints on the core emission (since they can separate the core
emission from knots of jet emission even within the innermost parsec
of the source).  High resolution optical (HST) and X-ray ({\it ROSAT}
HRI) measurements are likely to be free of galactic contamination but
may still possess a significant contribution from the inner jet.
Sub-mm and far-IR studies provide uninteresting limits on the core
flux: the comparatively poor spatial resolution leads to severe
contamination from galactic emission (predominantly thermal emission
by dust).

\subsection{The 3C~273 data}

For comparison with M87, the open squares on Fig.~2 show data for the
radio-loud quasar 3C~273.  These data are from the compilation of
Lichti et al. (1995) and are simultaneous or near-simultaneous.  Much
of this emission is likely to originate from the jet and be
relativistically beamed.  However, the fact that we see a big blue
bump and optical/UV emission lines from 3C~273 implies that a
significant part of the optical/UV emission is unbeamed and likely to
originate from the accretion flow.

\section{The advection-dominated model}

It is well known that sub-Eddington accretion can proceed via a thin
accretion disk (see review by Pringle 1981 and references therein).
Such disks are characterised by being radiatively efficient so that
the energy generated by viscous dissipation is radiated locally.  As a
consequence, such disks are cold in the sense that the gas temperature
is significantly below the virial temperature.  However, for
sufficiently low mass accretion rates ($\mdot<\mdot_{\rm crit}\approx
0.3\alpha^2$, where $\alpha$ is the standard disk viscosity parameter)
there is another stable mode of accretion (see Narayan
\& Yi 1995 and references therein).  In this second mode, the
accretion flow is very tenuous and, hence, a poor radiator.  The
energy generated by viscous dissipation can no longer be locally
radiated -- a large fraction of this energy is advected inwards in the
accretion flow as thermal energy and, eventually, passes through the
event horizon.  These are known as advection-dominated disks (ADDs).

For convenience, we rescale the radial co-ordinate and define $r$ by
\begin{equation}
r=\frac{R}{R_{\rm Sch}},
\end{equation}
where $R$ is the radial coordinate and $R_{\rm Sch}$ is the
Schwarzschild radius of the hole.  It is an important feature of ADDs
that, in the region $r<1000$, the ions and electrons do not possess
the same temperature.  The ions attain essentially the virial
temperature.  They couple weakly via Coulomb interactions to the
electrons which, due to various cooling mechanisms, possess a
significantly lower temperature.

By assuming that the system is undergoing advection-dominated
accretion, we can predict the radio to X-ray spectrum of the accretion
flow.  Since the gas is optically-thin, the emission in different
regions of the spectrum is determined by synchro-cyclotron,
bremsstrahlung and inverse Compton processes. The amount of emission
from these different processes and the shape of the spectrum can be
determined as a function of the model variables: the viscosity
parameter, $\alpha$, the ratio of magnetic to total pressure, $\beta$,
the mass of the central black hole, $M$, and the accretion rate,
$\dot{m}$.  For the moment, we take $\alpha=0.3$ and $\beta=0.5$
(i.e. magnetic pressure in equipartition with gas pressure), although
see discussion in Section~5.  The electron temperature at a given
point in the ADD, $T_{\rm e}$, can then be determined
self-consistently for a given $\dot{m}$ and $M$ by balancing the
heating of the electrons by the ions against the various radiative
cooling mechanisms.  Within $r<1000$, it is found that $T_{\rm
e}\approx 2\times 10^9\K$.  To determine the observed spectrum, we
must integrated the emission over the volume and take account of
self-absorption effects.  We have taken the inner radius of the disk
to correspond with the innermost stable orbit around a Schwarzschild
black hole, $r_{in}=3$, and the outer radius to be $r_{out}=10^3$.
The spectrum is rather insensitive to the choice of $r_{\rm out}$
since most of the radiation originates within $r_{out}$.  Details of
the model, which is based on that of Narayan \& Yi (1995), can be
found in Di~Matteo \& Fabian (1996).

In Fig.~2 we show the spectrum of the advection-dominated disk for
$M=3\times 10^9\Msun$ and $\dot{m}=10^{-3.5},10^{-3.0}$ and
$10^{-2.5}$.  The peak in the radio band is due to synchro-cyclotron
emission by the thermal electrons in the magnetic field of the
plasma. The X-ray peak is due to thermal bremsstrahlung emission.  The
power-law emission extending through the optical band is due to
Comptonization of the synchro-cyclotron emission: more detailed
calculations show this emission to be comprised of individual peaks
corresponding to different orders of Compton scattering.  The
positions at which the synchrotron and bremsstrahlung peaks occur and
their relative heights depend on the parameters of the model.  The
synchrotron radiation is self-absorbed and gives a black body
spectrum, up to a critical frequency, $\nu_c$.  Above the peak
frequency the spectrum reproduces the exponential decay in the
emission expected from thermal plasma (Mahadevan \& Narayan 1996).
The bremsstrahlung peak occurs at the thermal frequency $\nu\sim
k_{\rm B}T_e/h$.

 \begin{figure*}
 \centerline{\psfig{figure=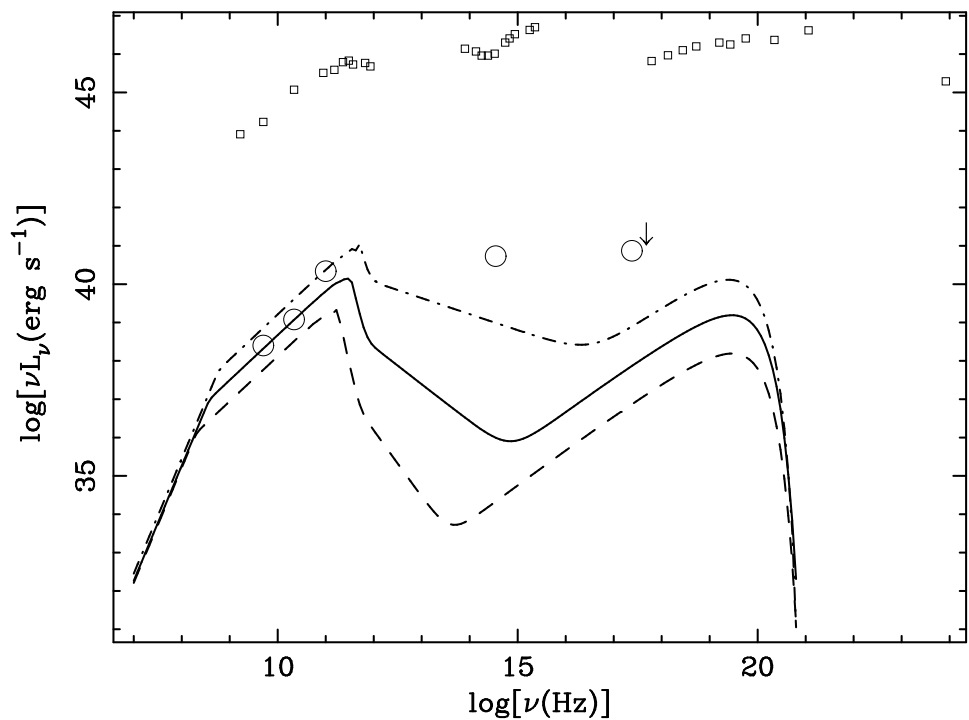,width=0.8\textwidth,angle=270}}
 \caption{Spectra of M87 calculated with an advection-dominated flow
 extending from $r_{min}=3$ to $r_{max}=1000$. The parameters are
 $m=3\times10^9$, $\alpha=0.3$, $\beta=0.5$. Three models are shown:
 (i) $\dot{m}=10^{-3.5}$-dashed line, (ii) $\dot{m}=10^{-3}$-solid
 line, (iii) $\dot{m}=10^{-2.5}$-dot-dashed line.  The circles and
 squares represent various measurements of the spectrum of M87 and
 3C~273, respectively, taken from different the references explained
 in the text.  For M87, a distance of $16\Mpc$ is assumed, whereas for
 3C~273 we have taken a Hubble constant of $50\kmps\Mpc^{-1}$ and a
 deceleration parameter of $q_{\rm 0}=0.0$.  In our model, the
 spectrum of M87 would lie only 2 decades below that of 3C273 if it
 had a standard, rather than advection-dominated, disk.}
 \end{figure*}

\section{The comparison}

Figure~2 demonstrates the comparison between the core data and the
advection dominated disk model described above.  The accretion rate
for all three models shown is comparable (within an order of
magnitude) to the Bondi rate.  Each of these models represents a
physically plausible accretion rate.  The two lower model curves
($\mdot=10^{-3.5},10^{-3.0}$) are completely acceptable in the sense
that they do not exceed any observational bounds.  Furthermore, it can
be seen that a substantial portion of the VLBI core flux may originate
from an advection dominated disk. In particular two of the VLBI data
points seem to reproduce almost exactly the slope of self-absorbed
cyclo-synchrotron spectrum.  We note that radio observations of
early-type galaxy cores typically show rising or flat spectra with
very similar slope ($\approx 0.3$) (Slee et al. 1994), which is well
accounted for by the spectrum of an ADD (Fig.~2 and Mahadevan 1996).

The core as seen in the 100\,GHz VLBI data, the HST data and the {\it
ROSAT} HRI X-ray data requires some additional component.  The
synchrotron jet would be a candidate for this additional component
(provided the jet becomes self-absorbed at frequencies $\nu\approx
100$\,GHz or less).

Figure~2 shows the contrast between M87 and 3C~273.  3C~273 is
observed to be at least 5 orders of magnitude more luminous than M87
at all wavelengths.  The big blue bump in the spectrum of this quasar
is often interpreted to be thermal emission from a standard thin-disk.
Assuming a thin-disk efficiency of $\eta=0.1$, the inferred accretion
rate is $\Mdot\sim 50\Msun$.  Thus, it is possible that the mass
accretion rates in M87 and 3C~273 differ by only 2 orders of magnitude
despite the fact that the luminosities differ by 5--6 orders of
magnitude.

\section{Summary and discussion}

We conclude that accretion from the hot gas halo at rates comparable
with the Bondi rate is compatible with the low-luminosity of this core
provided the final stages of accretion (where most of the
gravitational energy is released) involve an advection dominated disk.
This is in complete accord with the suggestion of Fabian \& Rees
(1995).  An important test of this model will be the micro-arcsec
radio imaging such as is promised by the VLBI Space Observatory
Program (VSOP).  This will provide the capability to image the core of
M87 on scales comparable to the Schwarzschild radius of the black
hole.  Any advection-dominated disk should be directly revealed
as a synchrotron self-absorbed structure with a position angle that is
perpendicular to the jet axis.

In constructing our ADD model, we have assumed a large value of the
viscosity parameter, $\alpha=0.3$.  It should be stressed that the ADD
model remains compatible with the core data even for smaller values of
$\alpha$.  For $\alpha<0.3$, we have to postulate that the magnetic
field in the disk is below its equipartition value (i.e. $\beta>0.5$) in
order that an ADD with $\mdot\sim 10^{-3}$ does not violate the VLBI
limits.  Given the recent MHD disk simulations of Stone et al. (1996),
this is not an unreasonable relaxation of our basic assumptions.  In
fact, all that is required is that $\alpha$ is sufficiently large so
as to make the ADD solution accessible for the accretion rates present
in M87.  If $\mdot\sim 10^{-3}$, then we require $\alpha\approxgt 0.06$.

The consistency of this model with the data allows us to explore some
astrophysical implications of ADDs in giant elliptical galaxies.  Rees
et al. (1982) have argued that electromagnetic extraction from black
holes surrounded by ion-supported tori (which share many of the basic
physical properties of ADDs) can power relativistic jets in radio
galaxies. The jets can be collimated in the inner regions of the
ion-supported torus. If we apply the Blandford-Znajek (1977) mechanism
to M87, we predict that the amount of energy extracted from a Kerr
black hole is of the order of
\begin{equation}
L_{\rm EM}\approx10^{43}\left(\frac{a}{m}\right)^{2}B_2^2 M_{9}^2
\ergps,
\end{equation}
where $a<m$ is the usual angular momentum of the black hole (in
dimensionless units), $M=10^9M_9\Msun$ is the mass of the black hole
($M_{9}\approx 3$) and $B=10^2B_2$\,G is the magnetic field in the
vicinity of the hole.  From our model (assuming equipartition), we
determine $B$ to be
\begin{equation}
B_2=1.50 \left(\frac{r}{3}\right)^{-5/4} \left(\frac{\dot{m}}{10^{-3}}\right)^{1/2}. 
\end{equation}
This determination of $L_{\rm EM}$ is completely consistent with the
kinetic luminosity of M87 jet ($L_{\rm K}\sim 10^{43}\ergps$) obtained
by Reynolds et al. (1996).

ADDs may be relevant to the creation of a unified model for radio-loud
AGN.  We are suggesting that M87, a classic FR-I radio source,
possesses an ADD.  However, {\it ASCA} observations of broad iron
K$\alpha$ fluorescence lines in the two FR-II sources 3C~390.3
(Eracleous, Halpern \& Livio 1996) and 3C~109 (Allen et al. 1996)
strongly point to the presence of standard cold (thin) accretion disks
in these objects.  This leads to the speculation that the FR-I/FR-II
dichotomy is physically the dichotomy between advection-dominated and
standard disks: i.e. FR-I sources possess ADDs whereas FR-II sources
possess `standard' thin accretion disks.  In this sense, the accretion
disks in FR-II sources and Seyfert nuclei would be intrinsically
similar (Seyfert nuclei also display broad iron emission lines that
are believed to originate from the central regions of a thin accretion
disk.)  We note that a very similar unification scheme (in the context
of ion-supported tori versus thin accretion disks) has been discussed
by Begelman, Blandford \& Rees (1984) and Begelman (1985; 1986).  The
only difference is the increased observational evidence.

Extending this unified scheme to all AGN, the two fundamental classes
of AGN would be those with ADDs and those with thin accretion disks.
Within these two classes, some currently unknown physical difference
(possibly black hole spin parameter or galactic environment) would
create the dichotomy between radio-loud and radio-quiet objects.  It
has recently been suggested that narrow emission line galaxies (NELGs)
are radio-quiet nuclei possessing ADDs (Di Matteo \& Fabian 1996).
Within the context of our unified scheme, this leads us to identify
NELG and FR-I radio galaxies as basically the same objects, differing
only in the fact that FR-I objects possess radio-jets.  It is
interesting to note that broad optical emission lines are not observed
from either FR-I sources or `genuine' NELG\footnote{Some objects
classified as NELG are known to be obscured Seyfert nuclei.  The
obscuration usually reveals itself at other wavelengths (e.g. via the
associated X-ray absorption and/or IR bump resulting from reprocessing
by the obscuring dust.)  There are objects (which we called `genuine'
NELG) that have no evidence for obscuration but still do not show
broad optical emission lines.}.  Indeed, optical spectroscopy of the
core of M87 by the Faint Object Spectrograph (FOS) on {\it HST} fails
to reveal broad (FWHM$>2000\kmps$) H$\beta$ line emission (Harms et
al. 1994).  Thus, the presence of an ADD may preclude the formation of
a broad emission line region, possibly due to the lack of any strong
photoionizing continuum.

The most likely control parameter determining which mode of accretion
operates is the mass accretion rate $\mdot$.  The accretion could be
advection-dominated at low accretion rates ($\mdot<\mdot_{\rm crit}$)
and via a standard disk for larger accretion rates.  The smaller
radiative efficiency of ADDs provides a natural explanation for the
fact that FR-I sources are of lower power than FR-II sources.
However, the situation can be much more complex.  The mode of
accretion may depend on the thermal state of the inflowing material.
In an elliptical galaxy, accretion from a hot ISM may favour the ADD
mode (provided $\mdot$ is sufficiently low) simply because the
material does not have time to cool (see a detailed discussion of
this, in the context of ion-supported tori, by Begelman 1986).
Furthermore, the mode of accretion could be history dependent.  As
discussed by Fabian \& Crawford (1990), a thin disk is a prolific
radiator of optical/UV radiation, so that hot infalling material would
be Compton cooled by these soft photons and enter the accretion disk
pre-cooled.  By this mechanism, a thin disk could remain intact even
if the accretion rate were to drop into the regime in which an ADD is
allowed.  Some catastrophic event (such as a galaxy or sub-cluster
merger) would be required to make the transition from a thin disk to
an ADD.

Both Fabian \& Rees (1995) and Di Matteo \& Fabian (1996) have
suggested that ADDs are important for understanding the demise of
quasars since early cosmological epochs ($z\sim 2$).  A decrease in
the average mass accretion rate (associated with the galactic systems
becoming more relaxed and mergers/interactions becoming less
frequent), or an increase in the temperature of the accreting gas
(perhaps due to a decrease in Compton cooling following disruption of
a thin-disk) could lead quasars to undergo a transition and accrete
via ADD.  After the transition, the luminosity could be lower by
orders of magnitude, even though the mass accretion rate may only have
dropped by a relatively small fraction.  Such objects would thus be
undetectable at cosmological distances.  This suggestion has important
implications for estimates of the black hole mass.  Even quiescent
galaxies might contain black holes that are doubling their mass on
timescales of $\sim 10^{10}\yr$ (using the parameters of M87: note
that an M87-like object placed at cosmological distances would be
considered to be a weakly active object).  Shorter term behaviour of
AGN might also reflect changes in accretion mode.  It is known that
radio-loud objects are not steady over cosmological timescales but
undergo bursts of duration $10^6$--$10^8\yr$.  It is normally thought
that these bursts have to be caused by large changes in mass accretion
rate.  However, a burst could be caused by a relatively small increase
in $\mdot$ if that increase were sufficient to force an ADD to a
thin-disk transition.

\section*{Acknowledgements}

CSR and TDM acknowledge PPARC and Trinity College, Cambridge, for
support.  ACF thanks the Royal Society for support.  UH thanks the NRC
for support. CRC acknowledges partial support from NASA LTSA grant
NAGW 2681 through Smithsonian grant SV2-62002 and contract NAS8-38249.


\begin{thebibliography}{}
\bibitem{} Abramowicz M., Chen X., Kato S., Lasota J.~P., Regev O., 1995, ApJ, 438, L37
\bibitem{} Allen S.~W., Fabian A.~C., Idesawa E., Inoue H., Kii T., Otani C., 1996,  submitted.
\bibitem{} B{\"a}{\"a}th L.~B. et al., 1992, A\&A, 257, 31
\bibitem{} Begelman M.~C., Blandford R.~D., Rees M.~J., 1984, Rev. Mod. Phys., 56, 255
\bibitem{} Begelman M.~C., 1985, in Astrophysics of active galaxies and quasi-stellar objects, eds Miller J.~S., University Science Books, Mill Valley, P411
\bibitem{} Begelman M.~C., 1986, Nat, 322, 614
\bibitem{} Biretta J.~A., Stern C.~P., Harris D.~E., 1991, AJ, 101, 1632
\bibitem{} Blandford R.~D., Znajek R., 1977, MNRAS, 179, 433
\bibitem{} Bondi H., 1952, MNRAS, 112, 195
\bibitem{} Di Matteo T., Fabian A.~C., 1996, submitted
\bibitem{} Eracleous M., Halpern J.~P., Livio M., 1996, ApJ, 459, 89
\bibitem{} Fabian A.~C., Canizares C.~R., 1988, Nat, 333, 829
\bibitem{} Fabian A.~C., Crawford C.~S., 1990, MNRAS, 247, 439
\bibitem{} Fabian A.~C., Rees M.~J., 1995, MNRAS, 277, L55
\bibitem{} Ford H.~C. et al. 1995, ApJ, 1994, 435, L27
\bibitem{} Hansen C., Skinner G., Eyles C., Wilmore A. 1990, MNRAS, 242, 262
\bibitem{} Harms R.~J. et al., 1994, ApJ, 435, L35
\bibitem{} Lea S., Mushotzky R.,Holt S., 1982, ApJ, 262, 24
\bibitem{} Lichti G.~G. et al., 1995, A\&A, 298, 711
\bibitem{} Mahadevan R., Narayan R., 1996, ApJ, 465, 327
\bibitem{} Mahadevan R., 1996, preprint.
\bibitem{} Matsumoto H., Koyama K., Awaki H., Tomida H., Tsuru T., Mushotzky R., Hatsukade I., 1996, PASJ, 48, 201
\bibitem{} Mushotzky R.~F., Done C., Pounds K.~A., 1993, ARAA, 31, 717
\bibitem{} Narayan R., Yi I., 1995, ApJ, 452, 710
\bibitem{} Narayan R., Yi I., Mahadevan R., 1995, Nat, 374, 623
\bibitem{} Pauliny-Toth I.~I.~K., Preuss E., Witzel A., Graham D., Kellerman K.~I., Ronnang B., 1981, AJ, 86, 371
\bibitem{} Pringle J.~E., 1981, ARAA, 19, 137
\bibitem{} Rees M.~J., 1982, in Riegler G., Blandford R., eds, The Galactic Center. Am. Inst. Phys., New York, p. 166.
\bibitem{} Rees M.~J., Begelman M.~C., Blandford R.~D., Phinney E.~S., 1982, Nat., 295, 17 
\bibitem{} Reid M.~J., Biretta J.~A., Junor W., Muxlow T.~W.~B., Spencer R.~E., 1989, ApJ, 336, 112
\bibitem{} Reynolds C.~S., Fabian A.~C., Celotti C., Rees M.~J., 1996, MNRAS, in press
\bibitem{} Sadler E.~M., Jenkins C.~R., Kotanji C.~G., 1989, MNRAS, 240, 591
\bibitem{} Slee O.~B., Sadler E.~M., Reynolds J.~E., Ekers R.~D., 1994, MNRAS, 269, 928
\bibitem{} Spencer R.~E., Junor W., 1986, Nat., 321, 753
\bibitem{} Stark A.~A., Gammie C.~F., Wilson R.~W., Bally J., Linke R.~A., Heiles C., Hurwitz M., 1992, ApJS, 79, 77
\bibitem{} Stone J.~M., Hawley J.~F., Gammie C.~F., Balbus S.~A., 1996, ApJ, 463, 656
\bibitem{} Takano S., Koyama K., 1991, PASJ, 43, 1
\bibitem{} Wrobel J.~M., Heeschen D.~S., 1991, AJ, 101, 148
\end{thebibliography}
\end{document}